\documentstyle[emulateapj,epsfig,apjfonts]{article}

\lefthead{L\'opez-Mart\'\i n et al.}
\righthead{Photoevaporating flows from  the cometary knots in the Helix nebula (NGC 7293)}

\def \yskip{\penalty-50\vskip3pt plus 3pt minus 2pt}
\def \pp{\par \yskip \noindent \hangindent .4in \hangafter 1}
\def \abc#1#2#3#4 {\pp#1, {\sl#2}, {\bf#3}, #4}
\def \blank {\lower 5pt\hbox to 0.75in{\hrulefill}}

\newfont{\rten}{cmr10}

\begin{document}

\title{Photoevaporating flows from  the cometary knots in the Helix nebula (NGC 7293)
\footnote{Partially based on observations made with the NASA/ESA
Hubble Space Telescope, obtained from the data archive at the
Space Telescope Science Institute. }}

\author{L. L\'opez-Mart\'\i n$^1$, A. C. Raga\altaffilmark{1},
G. Mellema\altaffilmark{2}, W. J. Henney\altaffilmark{3}, J. Cant\'o$^1$}
\altaffiltext{1}{Instituto de Astronom\'\i a,
UNAM, Apdo. Postal 70-264, 04510 M\'exico, D. F., M\'exico}
\altaffiltext{2}{Stockholm Observatory, S-133 36 Saltsj\"obaden, Sweden}
\altaffiltext{3}{Instituto de Astronom\'\i a, UNAM,
J. J. Tablada 1006, Colonia Lomas de Santa Mar\'\i a,
58090 Morelia, Michoac\'an, M\'exico}

\begin{abstract}
We explain the H$\alpha$ emission of the cometary knots in the 
Helix Nebula (NGC 7293) with an analytical model that describes the 
emission of the head of the globules as a photoevaporated flow produced 
by the incident ionizing radiation of the central star.
 We compare these models with the H$\alpha$ emission obtained from 
the HST ({\it Hubble Space Telescope}) archival images of the Helix Nebula.
  From a comparison of the H$\alpha$ emission with the
 predictions of  the analytical model we obtain a rate of ionizing  photons
 from the central star of about $\rm 5 \times 10^ {45} \ s^{-1}$, which is 
consistent with estimates based on the total H$\beta$ flux of the nebula. 
We also model the tails of the cometary knots as a photoevaporated wind 
from a neutral shadow region produced by the  diffuse ionizing photon
 field of the nebula. A comparison with the HST  images allows us to 
obtain a direct determination of the value of the diffuse ionizing flux. 
We compare the ratio of diffuse to direct stellar flux as a function 
of radius inside an HII region with those obtained from the
observational data through the 
analytical tail and head wind model. The agreement of this model with the
values determined from the observations of the knots is excellent.

\end{abstract}



\section{Introduction}

The small scale structure of the Planetary Nebula (PN) known as the
Helix Nebula (NGC 7293, PK 36-57$^\circ$1) is characterized by many thousands
of small knots. These knots (which were first reported  by 
Vorontsov--Velyaminov 1968) have a cometary shape, with their tails pointing away from
the central source. Groundbased work (Meaburn et al. 1992, 1996, 1998) and Wide Field and Planetary Camera HST observations (O'Dell \& Handron 1996; O'Dell \& Burkert 1997; Burkert \& O'Dell 1998) revealed a
multitude of spatially resolved knots, about 3500, with highly symmetric appearance.
The knots have also been detected in CO (Huggins et al.~1992) and,
at least outside the main nebula, in C~I (Young et
al.~1997). 

This PN is one of the closest to us with a parallax distance of 213~pc
(Harris et al. 1997). Other methods have been applied to determine the distance to this PN,
giving  values ranging from  120 to 400 ~pc. Optical images show a complicated morphology
characterized by a helical structure in H$\alpha$ and [N~II] and a more
elliptical shape in [O~III]. The deprojection of these images has
proven to be difficult, with suggestions of both elliptical/toroidal
shapes and bipolar ones (Meaburn \& White~1982) being made. O'Dell (1998) has  suggested
that the ionized region is a disk.

The central star is very hot and has a low luminosity, indicating 
that it is well down the cooling track of its post-AGB evolution.
The temperature of the central star of the Helix has been measured by
the H$\beta$ Zanstra method to be $\simeq 1.2 \times 10^5\,
\rm {K}$ (G\'orny, Stasi\'nska \& Tylenda 1997). 

The high effective temperature combined with the fairly low luminosity
and the large size of the PN , about 0.5~pc  in  diameter at a  distance of
213~pc,  leads to a complicated ionization structure with
co-existing regions of high and low ionization character (O'Dell 1998; 
Henry, Kwitter \& Dufour 1999). The low luminosity of the star also implies 
a very low density wind, making it virtually undetectable. However, with 
its expected high velocity ($v_\infty \simeq 6000$~km~s$^{-1}$ if radiation 
driven) it could still have a dynamical effect on the PN structure.

The measured densities in the nebula are low, about  60~cm$^{-3}$ (O'Dell 1998),
in contrast with the densities derived for the neutral/molecular globules, 
$n_{\rm knot}\sim 10^6$~cm$^{-3}$ (O'Dell \& Handron~1996). This leads to
the curious result
that a substantial fraction of the mass of the PN is in the form of
these neutral globules. The aspherical shape of the PN is also found 
in the distribution of the knots and in the large scale CO emission. Meaburn et
al.~(1998) have shown that the knots follow the same velocity
distribution as the CO, although at a lower typical expansion
velocity. Young et
 al.~(1999) found the deprojected expansion velocities of the knots and
the CO ring to be 19 and 29 $\rm km~s^{-1}$, respectively. 

Some other authors have treated the photoevaporation of clumps in 
an ionizing radiation field. Bertoldi \& McKee (1990) developed an 
approximate analytic theory of the evolution of a photoevaporating 
cloud exposed to the ionizing radiation of a newly formed star, finding
an equilibrium cometary cloud configuration. Johnstone, Hollenbach
\& Bally (1998) modelled the photoevaporation of dense clumps of gas by an 
external source of ultraviolet radiation including thermal and dynamical
effects. Mellema et al. (1998) study the evolution of dense neutral clumps
located in the outer parts of a planetary nebula (like  the Helix
nebula).

\begin{figure*}
\begin{center}
\epsfig{file=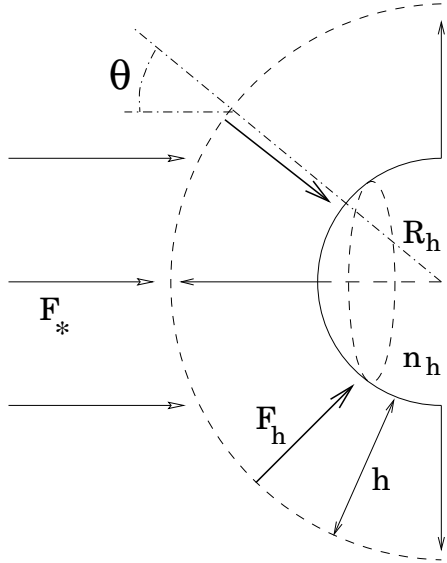,width=5.8cm}
\end{center}
\figcaption{Schematic diagram of the cometary head. $F_h$ is the incident flux at the surface
of the neutral knot, and $R_h$ and $n_h$ are the radius and the density of the neutral
knot (respectively).
\label{fig1}}
\end{figure*}

\section{Analytical  model for the photoevaporated flux of the cometary knots}

Let us discuss a simple, analytical model for cometary knots  embedded in a Planetary
Nebula (PN). First, we derive some expressions that relate the ionizing stellar flux
and the particle flux of the photoevaporated wind in a similar way to
Mellema et al. (1998)  to find the $\rm H\alpha$ emission for
the cometary heads. Cant\'o et al. (1998) proposed the idea of the tails as 
neutral shadow regions behind the clumps, and studied the complex 
time-evolution of the resulting flow. In this paper, we model the
tails behind the Helix clumps as a cylinder of neutral material being 
photoionized by the diffuse ionizing flux of the nebula.

\subsection{A model for the cometary head}

Let us  consider the problem of a hemispherical,  neutral knot  of radius $ R_h$ which is
being photoionized by the radiative field emitted by the central star of the ionized nebula.
A schematic diagram of this configuration is shown in Figure~\ref{fig1}.

A fraction of the stellar flux arrives at the knot surface, photoionizing the neutral
knot material and feeding a photoevaporated flow. The remaining ionizing photons
are absorbed in this photoevaporated flow. The neutral knot is being photoevaporated, but we consider a quasi-steady state in which the radius of the knot $R_h$ is almost constant with  time.  The incident  flux at the knot surface is~:

$$F_h(\theta)=n_h(\theta) ~v(R_h) \,, \eqno(1)$$

\noindent
where $F_h$ is the flux of ionizing photons incident normal to the knot surface (per unit
area and time), $n_h$  and $v(R_h)$ are the density and   the velocity of the  photoevaporated wind
at the knot surface . We note that the rate of photoionizations depends on the  angle of incidence $\theta$ (see Figure~\ref{fig1}), having a maximum value in the front of the cometary head.

In order to obtain the density profile, we consider the conservation of particles in
the photoevaporated wind~:

$$R^2n(R,\theta)v(R)=R_h^2n_h(\theta)v(R_h) \,,\eqno(2)$$

\noindent
where $R$ is the radius directed outwards from the knot.

\noindent
We assume that the ionization front is thin compared with the effective 
thickness $h$ of the ionized flow and that $h$ in turn is small compared 
with $R_h$. In this case, if all diffuse photons are absorbed ``on-the-spot'',
 the photoionization balance is given by 

$$F_{\star}~\cos(\theta) \simeq  F_h(\theta) + h~n_h^2(\theta)~\alpha_B  \,,\eqno(3)$$

\noindent
where $\alpha_B$ is the case B recombination coefficient of hydrogen, which 
is assumed to be constant with position. The flow thickness $h$ is 
defined by~:

$$h\equiv\omega~R_h\,,\eqno(4)$$

\noindent   where the parameter $\omega$ is defined through the relation~:

$$\omega~n_h^2~R_h\equiv \int_{R_h}^{\infty} n^2(r)~dr \,.\eqno(5)$$

\noindent 
This parameter $\omega$ depends on the density profile of the photoevaporated wind. For a D-critical ionization front $(v(R_h)=c_i$, where $c_i$ is the isothermal sound speed of the ionized gas), $\omega \simeq 0.1$ (Henney \& Arthur 1998).
From equations (1-5) we then obtain a relation between the ionizing flux
from the central star $F_{\star}$,  and the incident flux at  the knot surface  $F_h(\theta)$~:

$$F_{\star}~\cos(\theta) \simeq F_h(\theta)  + F_h^2(\theta)  ~ {\omega~R_h ~\alpha_B \over {c_i^2}}\,,\eqno(6)$$

\noindent

The term on the left is the ionizing stellar flux as  a function of the  angle
of incidence, the first term on the right is the  incident ionizing flux on 
the knot surface that produces the photoionizations of the neutral material 
and the second term  is the flux absorbed in the photoevaporated wind.  If we
 define~:

$$\xi_h \equiv {c_i^2 \over {R_h~\omega~\alpha_B}}\,,\eqno(7)$$
\noindent equation (6) can be written as ~:

$$F_{\star}~\cos(\theta)\simeq F_h(\theta) + {F_h ^2(\theta) \over \xi_h}\,,\eqno(8)$$

We can see from equation (8) that it is possible to find  the incident flux 
$F_h(\theta)$ as a function of the ionizing flux $F_{\star}$ solving the quadratic equation to obtain~:
 
$$F_h(\theta) \simeq {\xi_h \over 2} \left[\left({4~F_{\star}~\cos(\theta) \over \xi_h}
+ 1\right)^{1/2} - 1 \right]\,.\eqno(9)$$

\noindent We find two different limiting situations~:

If $\xi_h/\cos(\theta)<< F_{\star}$ we have the $''${\bf \it Recombination Dominated Regime} $''$
in which the fluxes are related by the expression~:

$$F_h(\theta) \approx \sqrt{\xi_h~F_{\star}~\cos(\theta)}\,,\eqno(10)$$

\noindent
In this case, an important fraction of the ionizing flux is absorbed in the photoevaporated wind and  only a small fraction of the stellar flux  arrives at the knot surface. The recombinations in the column beyond $R_h$ mainly balance the stellar flux.

 If $\xi_h/\cos(\theta) >> F_{\star}$ we have the $''${\bf \it Flux Dominated Regime} $''$, in which~:

 $$F_h(\theta)  \approx F_{\star}~\cos(\theta)\,,\eqno(11)$$

\noindent In this regime almost all the stellar flux arrives at  the knot surface and there is no absorption in the photoevaporated flow. The paticle flux at $R_h$ is roughly equal to the stellar flux.

In order to compute the total rate  of $\rm H{\alpha}$ photons emitted by 
the head of a  cometary knot, assuming $\omega \ll 1$ we have to evaluate 
the integral~:

$$S_{\rm H\alpha} \simeq  \int_0^{\pi/2} 2~\pi~R_h^2~ \alpha_{\rm H\alpha}~n^2_h(\theta)~\sin(\theta)~\omega ~ R_h ~ d\theta \,,\eqno(12)$$

\noindent
where  $\alpha_{\rm H{\alpha}} $ is  the   effective $\rm {H{\alpha}}$ recombination coefficient.

Taking into account the angular dependence of the density profile  we can integrate (12) to obtain the $\rm H\alpha$ emission of the cometary heads~:

$$S_{\rm {H\alpha}} \simeq \pi R_h^2 {\alpha_{\rm H\alpha} \over \alpha_B}F_{\star}\left\{1+ {\xi_h \over F_{\star}} - {\xi_h^2 \over {6F_{\star}^2}}\left[ \left(1+{4F_{\star} \over \xi_h}\right)^{3/2}-1\right]\right\}\,.\eqno(13)$$

\noindent
The term in curly brackets is the fraction of incident ionizing photons 
that are absorbed in the photoevaporating flow before reaching the ionization 
front. It is only this fraction of the incident photons that are reprocessed 
into H$\alpha$ radiation. If we know the size of the knot $R_h$ and the 
H$\alpha$ emission $S_{\rm H{\alpha}}$ we can use equation (13) to estimate
 the stellar flux.

\begin{figure*}
\begin{center}
\psfig{figure=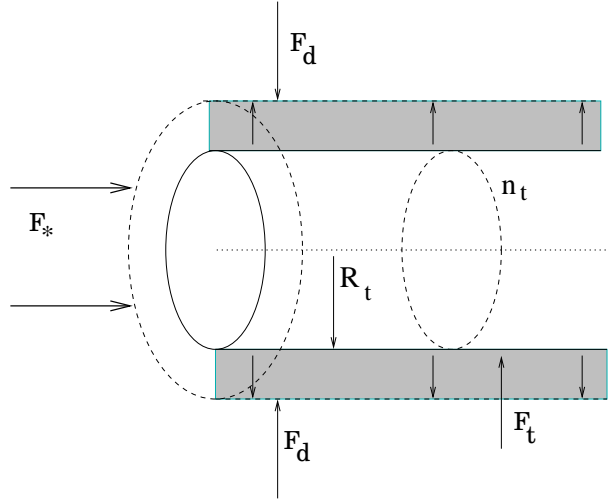,width=7.99cm}
\end{center}
\figcaption{Schematic diagram of the cometary tail. $F_t$ is the incident flux at
the surface of the neutral cylinder, $F_d$ is the diffuse flux (produced by the
surrounding nebula), and $R_t$ and $n_t$ are the radius and the density of the neutral tail
(respectively).
\label{fig2}}
\end{figure*}

\subsection{A model for the cometary tail}

For the cometary tail, we consider the problem of a cylinder of neutral material
behind the cometary head  being photoionized by the diffuse flux of the nebula. A schematic
diagram of this configuration is shown in Figure~\ref{fig2}. If we assume that the radius of the
cometary head is much smaller than the distance to the source  we can consider the shadow
region to have polar symmetry. This cylindrical shadow does not receive direct stellar radiation.
We therefore have~:

$$F_{d}=F_t+\omega~R_t~n_t^2~\alpha_B  \,,\eqno(14)$$
\noindent
where $F_d$ is the diffuse ionizing flux of the surroundings,  $F_t$ is the incident flux at the tail surface and the second term on the right represents the absorptions in the photoevaporated wind of the cometary tail. 

\noindent The incident flux at the neutral surface of the tail is related to the density $n_t$  at the base of the cylindrical wind through~:

$$F_t = n_t ~c_i \,,\eqno(15)$$

\noindent Defining  the parameter~:
$$\xi_t \equiv{{c_i^2} \over {\omega~\alpha_B~R_t}} \,,\eqno(16)$$

\noindent and subsituting equations (15-16) in (14) one obtains~:

$$F_d=F_t+{F_t^2 \over \xi_t} \,.\eqno(17)$$

\noindent 
In order to obtain the total number of $\rm {H{\alpha}}$ photons emitted by the tails of the
cometary knots, we assume that the H${\alpha}$ emitting region
has a cylindrical section with a thickness  $h\equiv\omega~R_t$. The number of H$\alpha$
photons emitted per unit time and per unit length of this cylinder is~:

$${S_{\rm H{\alpha}} \over {\Delta l}} = 2~\pi~R_t~\omega~R_t~n^2_t~\alpha_{\rm H{\alpha}}
\,,\eqno(18)$$
\noindent and as a function of the parameter $\xi_t$ the $\rm H\alpha$ emission is~:

$${S_{\rm H{\alpha}} \over {\Delta l}} = 2~\pi~R_t~ {\alpha_{\rm H{\alpha}} \over
\alpha_B}~{F_t^2 \over \xi_t} \,,\eqno(19)$$

\noindent 
where $\Delta l$ is a unit of length along the cylindrical tail. If we combine equations (17) and (19) we can calculate the diffuse ionizing flux as a function of the H${\alpha}$ emission per unit length and the radius of the cylinder~:
$$F_d={{S_{\rm H{\alpha}} /{\Delta l}} \over {2~\pi~R_t}}~{{\alpha_B} \over {\alpha_{\rm H{\alpha}}}}+ \left[{{S_{\rm H{\alpha}} /{\Delta l}} \over {2~\pi~R_t}}~{{\alpha_B} \over {\alpha_{\rm H{\alpha}}}}~\xi_t\right]^{1/2} \,.\eqno(20)$$

\subsection{Diffuse ionizing field inside an H II region}

As well as the radial ionizing radiation field from the Helix central
star, there will also be a diffuse ionizing radiation field,
principally due to ground level recombinations of hydrogen in the
nebula, plus smaller contributions from helium recombinations and the
scattering of stellar radiation by dust grains. For the purposes of
calculating the properties of the knot tails, the important quantity
is the lateral flux of the diffuse field, which is the flux incident
on the surface of an opaque, radially aligned, thin cylinder. 

In the ``on-the spot'' (OTS) approximation, in which all diffuse
ionizing photons are assumed to be reabsorbed by neutral H very close
to their point of emission, the diffuse flux, $F_{d}$, across
any opaque surface is related to the stellar flux, $F_*$, at the same
position by (Henney 2000) 

$$  \beta_{\rm OTS} \equiv \frac{F_{d}}{F_*} =
  \frac{\alpha_1}{4\alpha_{B} \kappa}  \,,\eqno(21)$$

\noindent 
where $ \alpha_1$ and $\alpha_B$ are, respectively, the H
recombination rates to the ground level and to all excited levels and
$\kappa = \bar{\sigma}_{d}/\bar{\sigma}_*$, where $\bar{\sigma}_{d}$ 
is the mean photoionization cross-section averaged over the diffuse 
ionizing spectrum and $\bar{\sigma}_*$ is the same quantity averaged 
over the stellar spectrum. Henney (2000) presents detailed
calculations, in which the OTS assumption is relaxed, for the lateral
diffuse flux in two simplified geometries: a classical filled-sphere
homogeneous Str\"omgren H II  region, and a hollow cavity H II region
with the ionized gas concentrated in a thin spherical shell. It is
found that in both cases the value of $\beta (=\frac{F_{d}}{F_*})$ 
is much smaller than
$\beta_{OTS}$ at small radii. It is because at small radi, $F_{\star}$
 gets huge ($r^{-2}$) but $F_d$, which is proportional to the
recombination rate times a path length, plateaus to a more or less
constant value.  In the thin-shell case,
$\beta$ remains at a low value throughout the interior of the
cavity, only becoming comparable to $\beta_{\rm OTS}$ at the
position of the shell. In the filled-sphere case, on the other hand,
for the relatively high values of $\kappa$ appropriate for the Helix
knots (see below), $\beta$ rises to $\simeq \beta_{\rm OTS}$ at
a fractional radius of $\simeq 0.5$ and is roughly constant
thereafter.

The stellar ionizing radiation field is much harder than the diffuse
field, while the photoionization cross-section declines rapidly with
frequency above the ionization threshold. Hence, the ratio of mean
cross-sections, $\kappa$, for the diffuse and stellar fields is
larger than unity. Including the H and He$^0$ recombination spectrum
gives $\bar{\sigma}_{d} \simeq 0.75 \sigma_0$, where
$\sigma_0$ is the threshold cross-section and $\rm {He/H} =
0.13$ (O'Dell 1998; Henry, Kwitter \& Dufour 1999) has been assumed.

The temperature of the central star of the Helix has been measured
by the H\(\beta\) Zanstra method to be \(\simeq 1.2 \times 10^5\,
\mathrm{K}\) (G\'orny, Stasi\'nska \& Tylenda 1997).
A further
constraint on the stellar spectrum is the observation (O'Dell 1998)
that the He$^{++}$ zone in the nebula is roughly half the radius of
the He$^{+}$ zone, implying that roughly 10$\%$ of the ionizing
photons have frequencies higher than the He$^+$ ionization limit.
Model atmospheres for compact hot stars (Rauch 1997) that are
consistent with these constraints have photon spectra $L_\nu/h\nu$
that are flat or rising between the H and He$^+$ ionization limits,
implying a value of $\bar{\sigma}_* \simeq 0.17 \sigma_0$. 

Hence, $\kappa \simeq 4.5$ is appropriate for the Helix knots, which
implies $\beta_{\rm OTS} \simeq 0.033$ for an assumed electron
temperature of $10^4\,{K}$. The true situation in the Helix
nebula is probably intermediate between the filled-sphere and the
hollow-shell case. Although O'Dell (1998) finds that the electron
density is roughly constant with radius throughout the nebula, the
central high-ionization ``hole'' has a higher temperature and hence a
lower emissivity of diffuse ionizing photons. Furthermore, the
geometry is more disk-like than spherical, which would tend to reduce
the intensity of the diffuse field. 

\begin{figure*}
\begin{center}
\epsfig{file=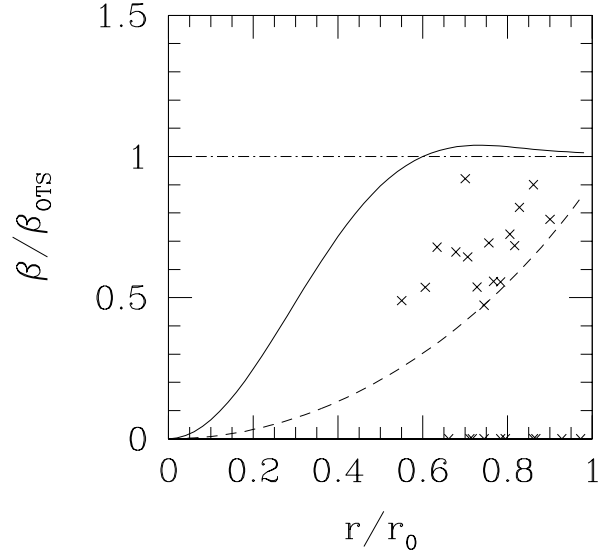,width=7.8cm}
\end{center}
\figcaption{Comparison of the Henney (2000) models to the observed values of  the diffuse to direct flux ratio (crosses) as a function of the distance from the ionization front. The upper horizontal line is the OTS value and the solid line and dashed line give the variation of $\beta$ across the nebula for the filled-sphere and hollow-shell case, respectively.
\label{fig3}}
\end{figure*}

The radial dependence of $\beta$ in the two limiting cases is shown in 
Figure~\ref{fig3}. This is compared with the observational data in section 3.3. 

\section{Properties of the cometary knots}

We have used the HST images from the data archive at the Space Telescope
Science Institute to obtain the H$\alpha$ emission from the cometary
knots in the Helix Nebula. These images are flux-calibrated using the 
coefficients given in O'Dell \& Doi 1999. The f656n and f658n filters was
 selected to isolated the  H$\alpha$ and [NII] emission at 658.4 nm.
The [NII] emission is not much stronger than  H$\alpha$, and thus it does
not produce an important contamination of the  H$\alpha$ observations (less 
than $\approx 15 \%$).

Several methods have been 
used in order to obtain the distance to this PN (Cahn \& Kaler 1971;
 Daub 1982; Cahn, 
Kaler \& Stanghellini 1992; Harris et al 1997), leading to distances 
ranging from 120 to 400 pc. We use the value of 213~pc determined by 
Harris et al.~(1997) from trigonometric parallax.

The comparison between the models and the observations is done for 26 
knots, chosen for being relatively well isolated, and covering a range 
of distances to the central star.

\subsection{Sizes of the knots}

In order to determine the sizes of the knots, we carry out aperture 
photometry with a series of circular diaphragms (centred on the knots) 
of increasing angular radius. If one plots the flux within the aperture 
versus radius, for large enough radii the flux has to increase
as the square of the radius (due to the presence of the bright, surrounding 
nebular environment). From a logarithmic plot it is then straightforward to
determine the radius $R_{ext}=R_h+\omega R_h$ (see equations 4-5 and Figure~\ref{fig1}) at which the quadratic flux vs. radius dependence first appears. 
In this way one then determines the value of the radius $R_h$ of the neutral
 clump.

\begin{figure*}
\begin{center}
\psfig{figure=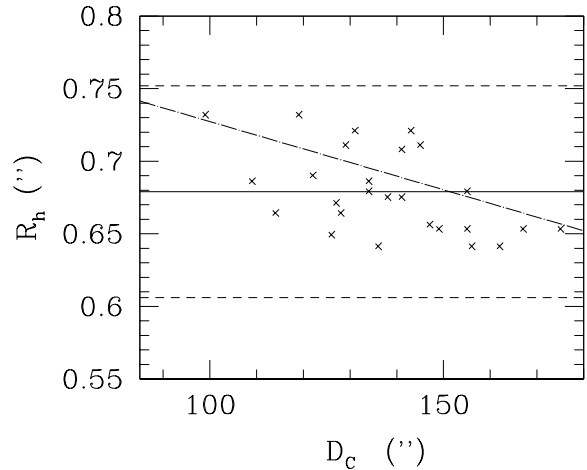,width=8.8cm}
\end{center}
\figcaption{The angular radii of the knots as a function of projected angular distance
to the central star. The crosses indicate the angular sizes determined for individual knots.
The horizontal, continuous line indicates the mean value of  knot radius,
and the dashed horizontal lines give the value of the dispersion from the mean value.
The oblique dotted line gives the best fit to the observed data.
\label{fig4}}
\end{figure*}

Applying this method to the 26 chosen knots we find that there is no  
correlation between the size of the knots and their distance to the central 
star (see Figure~\ref{fig4}). However, this could be a result of the fact that we only
 have a few knots, and that these knots are not necessarily representative 
of the true distribution of knot sizes (as they were chosen so as to be
well isolated, see above). We calculate a mean radius for the knots of 
$<R_h> = (0.68 \pm 0.07) ~''$.

\goodbreak
\subsection{H$\alpha$ intensities of the cometary heads}

Subtracting the background from the fluxes determined with the
aperture photometry, we obtain the H$\alpha$ fluxes emitted by the
knots. One can then use equation (13) to calculate the stellar ionizing 
flux at the position of the successive knots as a function of distance 
from the central star.

\def\pz{\phantom{0}}
\begin{table*}
 \centering
  \caption{Location and fluxes of the cometary knots}
  \begin{tabular}{@{}cccccccc@{}}
\hline
 $\alpha_{2000}$\footnote{System proposed by O'Dell \& Burkert idnetifying a knot by its right ascension in units of 0$^s$.1 (column 1) and by its declination in units of 1$''$ (column 2)} & $\delta_{2000}$$^a$  & $R_h ~('')$ & $S_{H\alpha}$\footnote{H$\alpha$ production rate of the cometary head} (s$^{-1}$) & $S_{H\alpha}$\footnote{H$\alpha$ production rate of the cometary tail} (cm$^{-1}$~s$^{-1}$) & $F_*$ ~(cm$^{-2}$~s$^{-1}$) & $F_d$ ~(cm$^{-2}$~s$^{-1}$) & $F_d/F_{\star}$ \\
\hline
429 & -860 & 0.73 & ${10.84\times 10^{39}}$ & ${2.96\times 10^{22}}$ & ${1.02\times 10^{10}}$ & ${1.67\times 10^{8}}$ & $1.63\times 10^{-2}$ \\
433 & -853 & 0.69 & ${5.16\times 10^{39}}$ & ${1.60\times 10^{22}}$  & ${7.21\times 10^{9}}$ & ${1.29\times 10^{8}}$ &  $1.79\times 10^{-2}$  \\
431 & -844 & 0.66 & ${4.67\times 10^{39}}$ &${2.53\times 10^{22}}$ & ${7.53\times 10^{9}}$ & ${1.70\times 10^{8}}$  & $2.26\times 10^{-2}$\\ 
440 & -848 & 0.73 & ${4.30\times 10^{39}}$ &  -- & ${5.93\times 10^{9}}$ & --  & -- \\
452 & -901 & 0.69 & ${3.44\times 10^{39}}$ & ${1.54\times 10^{22}}$ & ${5.74\times 10^{9}}$ & ${1.27\times 10^{8}}$  & $2.20\times 10^{-2}$\\
428 & -827 & 0.65 & ${2.52\times 10^{39}}$ &  ${2.25\times 10^{22}}$ & ${5.31\times 10^{9}}$ & ${1.63\times 10^{8}}$  & $3.07\times 10^{-2}$\\
459 & -905 & 0.67 & ${3.36\times 10^{39}}$ & ${1.61\times 10^{22}}$ & ${6.22\times 10^{9}}$ & ${1.33\times 10^{8}}$ &  $2.15\times 10^{-2}$ \\
413 & -818 & 0.66 & ${3.39\times 10^{39}}$ & -- & ${6.27\times 10^{9}}$ & --  & --\\
425 & -822 & 0.71 & ${3.67\times 10^{39}}$ & -- & ${5.93\times 10^{9}}$ & --  & --\\
352 & -815 & 0.72 & ${7.05\times 10^{39}}$ &  ${2.51\times 10^{22}}$ & ${8.68\times 10^{9}}$ &  ${1.56\times 10^{8}}$  & $1.79\times 10^{-2}$\\
410 & -808 & 0.68 & ${3.79\times 10^{39}}$ & -- & ${6.67\times 10^{9}}$ & --  & --\\
474 & -931 & 0.69 & ${6.11\times 10^{39}}$ &   ${1.52\times 10^{22}}$     & ${7.99\times 10^{9}}$ & ${1.26\times 10^{8}}$ & $1.58\times 10^{-2}$\\
473 & -919 & 0.64 & ${2.55\times 10^{39}}$ & ${1.53\times 10^{22}}$ & ${5.89\times 10^{9}}$ & ${1.36\times 10^{8}}$ & $2.31\times 10^{-2}$\\
378 & -800 & 0.67 & ${4.26\times 10^{39}}$ & ${1.58\times 10^{22}}$ & ${7.13\times 10^{9}}$ & ${1.32\times 10^{8}}$ & $1.86\times 10^{-2}$ \\
354 & -804 & 0.71 & ${5.14\times 10^{39}}$ & ${1.81\times 10^{22}}$ & ${7.21\times 10^{9}}$ & ${1.33\times 10^{8}}$ & $1.85\times 10^{-2}$ \\
465 & -853 & 0.67 & ${2.35\times 10^{39}}$ & -- & ${5.10\times 10^{9}}$ & --   & --\\
480 & -925 & 0.72 & ${2.45\times 10^{39}}$ & -- & ${4.73\times 10^{9}}$ & --   & --\\
351 & -802 & 0.71 & ${5.32\times 10^{39}}$ & ${3.16\times 10^{22}}$ & ${7.35\times 10^{9}}$ & ${1.78\times 10^{8}}$ & $2.41\times 10^{-2}$ \\
398 & -752 & 0.66 & ${3.22\times 10^{39}}$ & ${1.68\times 10^{22}}$ & ${6.06\times 10^{9}}$& ${1.38\times 10^{8}}$ & $2.28\times 10^{-2}$  \\
386 & -750 & 0.65 & ${3.31\times 10^{39}}$ &  ${1.78\times 10^{22}}$  & ${6.17\times 10^{9}}$ & ${1.68\times 10^{8}}$  & $2.73\times 10^{-2}$\\
360 & -751 & 0.68 & ${2.79\times 10^{39}}$ & ${2.62\times 10^{22}}$ & ${5.61\times 10^{9}}$& ${1.68\times 10^{8}}$ & $3.00\times 10^{-2}$ \\
352 & -750 & 0.65 & ${3.94\times 10^{39}}$ & -- & ${6.82\times 10^{9}}$ & --  & --\\
389 & -742 & 0.64 & ${2.74\times 10^{39}}$ & -- & ${6.11\times 10^{9}}$ & --  & --\\
363 & -740 & 0.64 & ${3.19\times 10^{39}}$ & ${2.45\times 10^{22}}$ & ${6.67\times 10^{9}}$ & ${1.73\times 10^{8}}$ & $2.59\times 10^{-2}$ \\
494 & -911 & 0.65 & ${2.77\times 10^{39}}$ & -- & ${5.55\times 10^{9}}$ & --  & --\\
372 & -725 & 0.65 & ${3.19\times 10^{39}}$ & -- & ${6.06\times 10^{9}}$ & --  & --\\
\hline
\end{tabular}
\end{table*}

\begin{figure*}
\begin{center}
\psfig{figure=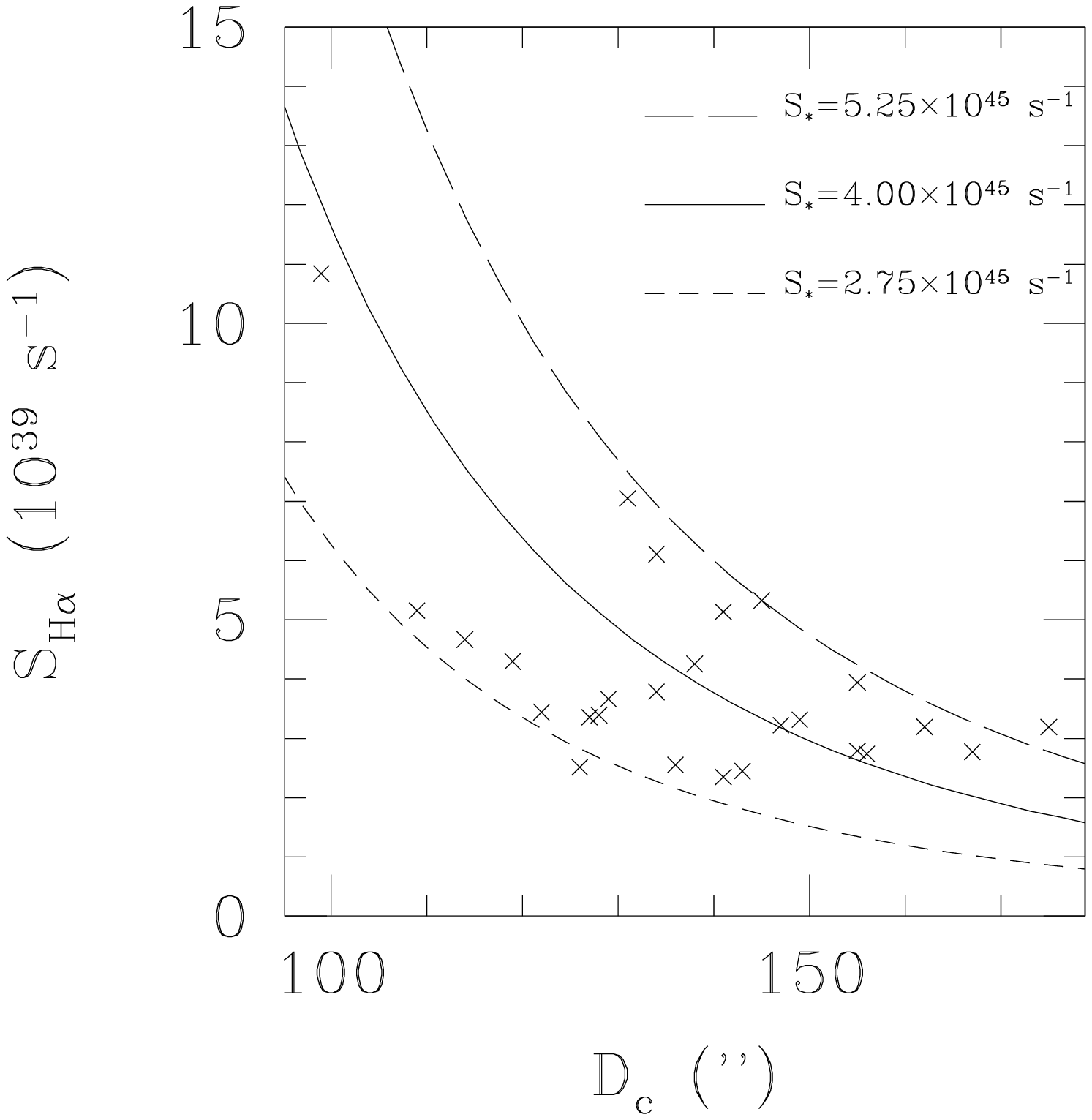,width=9.8cm}
\end{center}
\figcaption{Model fits to the H$\alpha$ photon rates observed for the knots (as a function
of distance from the central star) obtained for different stellar ionizing photon rates.
The predicted curves should represent upper envelopes of the observed points.
\label{fig5}}
\end{figure*}

Column 6 of Table 1 gives the resulting H$\alpha$ photon production rates $S_{\rm H\alpha}$
of the knots,  using values for $\omega=0.1$ and $c_i = 10 \ \rm km~s^{-1}$ (therefore $\xi_h \simeq 1.79 \times 10^{10} \ \rm cm^{-2}~s^{-1}$ for a knot with the mean radius $<R_h> = 0.68~''$). Figure~\ref{fig5} shows $S_{\rm H\alpha}$ vs. $D_h$ (where
$D_h$ is the projected distance between the source and the clumps).

Figure~\ref{fig5} also shows the $S_{\rm H\alpha}$ vs. $D_h$ predicted from equation (13)
for different values of $S_*$ ($=4\pi {D_h}^2 F_*$). The predicted curves should
represent upper envelopes of the observed points, as the real distances between
the clumps and the source are larger than the observed, projected distances.
From a comparison of the predicted
curves with the values measured for the clumps, we see that the observations can
best be fitted with a stellar ionizing photon rate of

$$S_{\star} \approx 5 \times 10^{45} ~\rm s^{-1}\,.\eqno(22)$$

If the knots were distributed isotropically around the central star, then 
the median knot would have a true distance $\simeq 1.15$ times greater than
 its projected distance. However, the lower envelope of the observed knot 
brightness distribution suggests that all knots lie within $\pm 45^\circ$ 
of the plane of the sky, consistent with a ring-like spatial distribution.

 According to Osterbrock (1989) $S_{\star}$ can be derived from the $\rm
H\beta$ luminosity~:

$$S_{\star} = {\alpha_B \over \alpha_{\rm H\beta}^{\rm eff}} {L(H\beta)
\over {h \nu_{\rm H\beta}}}\,.\eqno(23)$$
\noindent 
If we use the value of the $\rm
H\beta$ luminosity of the Helix Nebula obtained by O'Dell (1998) we have
a value for the stellar ionizing photon rate of $5.25 \times 10^{45}
 ~\rm s^{-1}$. As we can see the match with our fit is excellent.

\subsection{H$\alpha$ intensities of the tails}

In order to obtain the H$\alpha$ emission from the tails we integrate 
the observed emission over a rectangular area covering the region in 
which the tails are clearly detected.
We also integrate the emission in two small adjacent rectangular areas 
in order to determine the intensity of the nebular
background, which we subtract from the emission of the box containing 
the cometary tail.

Knowing the H$\alpha$ emission we can then calculate the diffuse flux 
with equation (20) for the 26 chosen knots. In this way we calculate the 
diffuse flux for different distances to the central star (see column 7 
of Table 1).  As we  have also computed the direct stellar flux from the 
emission of the heads of the knots (section 3.2),
we can then calculate the diffuse-to-direct stellar flux ratios,
and compare these values with the ones predicted from the (Henney 2000)
models described in section 2.3. This comparison is shown in Figure~\ref{fig3}, where we see that the observed points fall into the region 
between the curves delimited by the two models, indicating that the case 
of the Helix nebula is intermediate between the homogeneus sphere and the
 ``thin shell'' cases.

\subsection{Knot masses and evaporation rates}

From the analysis it is also possible to derive knot masses and evaporation
rates. Under the assumption that the knots are accelerating and have an
exponential density profile, one can derive  their mass  by

$$  M=\left(1-{2\over \pi}\right){16F_{\rm c}(0) m c_{\rm i}\over \pi c_{\rm
     n}^2}R_{\rm c}^3\,,\eqno(24)$$
\noindent
see Mellema et al.~(1998), equation (32). In this equation $m$ is the average mass
 per atom or ion, for wich we take $1.3m_{\rm H}$, $R_{\rm c}$ is the radius 
of the clump, and $c_{\rm n}$ is the
 isothermal velocity of sound in the neutral gas for which we take 
1~km~s$^{-1}$ (corresponding to 150~K). Applying this to the measured knots 
gives an average mass of $9\times 10^{-6}$~M$_\odot$ with knot masses ranging 
from $6\times 10^{-6}$~M$_\odot$ to $1.8\times 10^{-5}$~M$_\odot$. In the 
process one has to calculate the fraction of the ionizing flux which actually
 reaches the heads of the knots, $F_c(0)/F_*$, which comes out to be about 70 
to 80\%.

The mass range found is very close to the one derived by O'Dell \& Handron
(1996) and Meaburn et al.~(1998) from dust extinction measurements  and
by Huggins et al.~(1992) from CO measurements, showing that the sizes and shapes
of the knots are entirely consistent with the photoevaporation model.

The model also allows an estimate of the current mass loss rate from the
heads of the knots, which is given by

$$  {{\rm d}M\over {\rm d}t}= -F_{\rm c}(0)m\pi R_{\rm c}^2\,,\eqno(25)$$
\noindent
see Mellema et al.~(1998), equation (36). This gives an average of $-2.2\times
10^{-9}$~M$_\odot$~yr$^{-1}$ (ranging between -4.1 and $-1.6\times
10^{-9}$~M$_\odot$~yr$^{-1}$).

\section{Discussion}

The results of the previous sections show that our photoevaporation model
is entirely successful in explaining the current situation of the cometary
knots in the Helix nebula. One could say that we are using the cometary knots
as `probes' for the direct and diffuse UV radiation in the nebula, under the
assumption that both the heads and tails are photoevaporating. The results
from the probes are consistent with the results found by other means.

The question which remains unanswered in this approach is the origin and
final fate of the knots. Starting with the their final fate, the
photoevaporation models give an estimate of the evaporation time. 
The evaporation time in the ``Flux Dominated Regime'' is $t_{\rm evap} 
\approx R_{\rm c} ~ c_{\rm i} / c_{\rm n}^2$, so that once a clump
start to shrink by evaporation, the timescale gets shorter and shorter.
The
results from Mellema et al.~(1998) show that the evaporation time of a knot
is given by

$$  t_{\rm evap}=t_*\left({5 \over 6}+{\ln(1+\eta)\over 6\eta}\right)\,,\eqno(26)$$
\noindent
with

$$  t_*={48 \over \pi}\left(1- {2\over \pi}\right){c_{\rm i} R_{\rm c} \over
 c_{\rm n}^2\pi}\,,\eqno(27)$$
\noindent
and
\noindent
$$  \eta={2\alpha_{\rm B}F_* R_{\rm c} \over 3\pi c_{\rm i}^2}\,,\eqno(28)$$
\noindent
which  applied to the measured knots gives an evaporation time of 1.1 to
$1.2\times 10^4$ ~years. The time for the clumps to move from close 
to the star to their present position is about $6\times 10^3$ ~years.
In the ``Flux Dominated Regime'' the evaporation time
is $t_{\rm evap} \approx R_{\rm c}c_{\rm i} c_{\rm n}^2$, so that
once a clump start to shrink by evaporation, the timescale gets shorter
and shorter. There is an uncertainty in this number since the
sound speed in the neutral gas is not well defined.  Probably, the knot has a
range of temperatures, depending for instance on how far the molecule
dissociating photons penetrate. The evaporation time is inversely
proportional to  the temperature, the values quoted  being valid for 150~K.
Huggins et al.~(1992) cite a temperature of 25~K for the CO gas in the
knots.

One consequence of the photoevaporation is that gas is fed into the region
surrounding the knots. Interestingly enough the Helix Nebula seems to be
special among PNe in that its `central cavity' is not empty but is at least
partly filled with high ionization gas producing detectable amounts of [OIII]
lines, see for example O'Dell (1998) and Henry et al.~(1999). The density of
this gas is derived to be around 50~cm$^{-3}$, perhaps less. We conjecture
that this material was injected into the cavity by the photoevaporating
clouds.  Taking a radius of 200$''$ for the cavity, and assuming a spherical
shape (probably an overestimate of the volume and hence of the mass), one
finds that the evaporation of 3000 to 5000 knots of about 10$^{-5}$~M$_\odot$
can supply this amount of gas. This equals the mass and current number of
knots in the Helix. Possibly the gas filling the central cavity is from
already evaporated knots, or from the current knots. In Mellema~et al.~(1998)
it was shown that 50\% of the mass of a photoevaporating clump is
lost during the first `collapse phase'. Since the current knots apparently
are in the `cometary phase' they might easily have already lost half of their
initial mass.

The question of the origin of the knots can be divided into two questions:
``Why do they have a cometary shape?'' and ``What physical process is
responsible for their existence?''. Three models exist to explain their
shapes. Dyson et al.~(1993) showed that collisions between a supersonic wind
and a clump produce short stumpy tails, and that only the interaction between
a subsonic wind and a subsonic flow from a clump can produce long, comet--like
tails. When applying this theorem to our model we run into the problem that
the shapes derived in Dyson et al.~(1993) are the shapes of the contact
discontinuity between the flow from the clump and the wind, which we do not
trace in our photoionization description. The dynamical models in
Mellema~et al.~(1998) contain both the photoionization processes and the
interaction of the photoevaporation flow with the environment. In those
models the shape of the contact discontinuity is far from comet--like, as
predicted by Dyson et al.~(1993); but at the same time the contact
discontinuity is not a region which produces a lot of emission, as the
ionization front outshines it by many factors. However, in that model the
environment does not exert any ram pressure on the clump and its
photoevaporation flow. If the ram pressure in the environment is high
enough it will overwhelm the photoevaporation flow, and dominate the
dynamics. Observations show that the environment of the cometary knots has a
density of about 50~cm$^{-3}$ and a temperature of perhaps 20,000~K. If the
environment is flowing subsonically, this implies pressures of order
$10^{-10}$~dyn~cm$^{-2}$. The photoevaporation pressure is given by
$2F_{\rm c}m c_{\rm i}$ (see Mellema et al.~1998), which gives values of
about $10^{-8}$~dyn~cm$^{-2}$. Clearly the environment will not be able to
overwhelm the photoevaporation flow from the knots.

We cannot rule out the possibility that some time in the past the cometary
knots were shaped into their cometary shape through a wind--clump
interaction. In fact, below we will argue for such a scenario to explain the
origin of the knots.

Cant{\'o} et al.~(1998) suggested that long tails can be formed behind clumps
opaque to ionizing UV photons. The region behind the clump does not see any
direct ionizing photons from the star, just the diffuse UV radiation field,
which in general leads to denser, partly neutral tails forming behind the
clumps. However, these tails will be initially the same density as the
environment, fill up and recombine, striving towards pressure equilibrium
with the ionized environment.  The tails of the Helix knots are overpressured
compared to their environment, a situation which does not occur in the
scenario put forward in Cant{\'o} et al.~(1998). The overpressured tails
suggest that they were formed under different circumstances from the ones
in which  they are now.

Burkert \& O'Dell (1998) proposed that the knots are shaped by Ly$\alpha$
photon radiation pressure. This was partly motivated by the fact that the
evaporation flows appear to have exponential brightness profiles. These
autors found that this mechanism does not work in the cometary knots of the Helix nebula.
 A similar model was proposed for the Proplyds in the Orion nebula (O'Dell
1998). Henney \& Arthur (1998) showed that the exponential brightness
profiles are {\it not} inconsistent with photoevaporation flows and that
the radiation pressure from Ly$\alpha$ is at least one order of magnitude too
low to be significant. Their arguments also hold for the Helix Nebula 
cometary knots and we will not repeat them here.  The conclusion is that 
none of the three models can explain the origin of the long cometary tails 
under the current circumstances.

When we consider the origin of the knots there are basically two options, the
knots are either primordial or the result of instabilities. Meaburn et
al.~(1998) and Young et al.~(1999) favour the primordial model, the main
arguments being that the CO gas outside of the main nebula has a very clumpy
distribution, and that the velocity distribution of the cometary knots
follows that of the main CO emission, albeit at a lower expansion
velocity. Within this model it is possible that the tails were shaped as the
knots were run over by the main part of nebula. During this phase they would
be surrounded and eroded by subsonically flowing gas, which according to the
Dyson et al.~(1993) model would produce long thin tails. Also, the formation
of the tails as ionization shadows behind the clumps as proposed by Cant{\'o}
et al.~(1998) would in this case produce long, dense tails. After the main
nebular shell has passed by, the knots find themselves fully exposed to the
ionizing stellar radiation  and the tails to the diffuse UV radiation field
and both start photoevaporating, which is the stage in which we see them now.

O'Dell \& Handron (1996) prefer the model in which the knots are formed as
Rayleigh--Taylor instabilities on the inside of the swept up main nebula,
 proposed originally by Capriotti~(1973). The
numerical model of R--T instabilities presented in figure~4 in O'Dell \&
Burkert~(1997) seems to support this model; it shows elongated structures
pointing radially to the centre of the nebula. However, the much higher
resolution models of Walder \& Folini (1998) show the R--T instabilities to
be much more chaotic. The fingers in their model need a lot of processing
before they look anything like the cometary knots, and one would also expect
 much more variation in knot properties if they were formed in such
instabilities. In all we prefer the first option of primoridal knots.

\section{Conclusions}
In this paper we propose an analytical model for the heads of the
cometary globules in the Helix Nebula, in which the emission is assumed to come
from a flow which is being photoevaporated from the surface of a neutral clump.
Using the H$\alpha$ emission obtained from HST images of the Helix Nebula
(extracted from the HST archive, see also O'Dell \& Handron 1996; O'Dell \& 
Burkert 1997) and the predictions from our analytic model, we can calculate 
the ionizing stellar photon flux at the positions of the knots. 

We find that the knot brightnesses are fully consistent with the ionizing 
photon luminosity of ${S_{\star}} = 5.25\times 10^{45} ~\rm s^{-1}$ for the
 central star, deduced from the total H$\beta$ flux of the nebula. 
This value, however, is somewhat uncertain as in the present work we have made
no attempt to substract the [N II] 6583 flux from  the f656n filter. Given the
 rather strong [N II] emission of some of the Helix Knots, our H$\alpha$
flux determinations could in some cases lie up to $\approx$30 
The angular size of the knots has a value of $<R_h> = (0.68 \pm 0.07) ~''$
(where $R_h$ is the radius of the neutral clump forming the knot, see section
 3.1). From our small sample of 26 knots, we find no correlation between $R_h$
 and distance from the central star (see Figure~\ref{fig4}). 

The photoevaporation model also predicts masses for the knots, which come out
fully consistent with observationally derived masses. Their evaporation
time is between several thousands to 10$^4$ years, dependent on the 
temperature of the neutral gas in the knots.

In this paper we also present a model for the cometary tails as a flow which is
being photoevaporated from a neutral, cylindrical ``shadow region'' behind the
neutral clumps which form the knots. This flow is photoevaporated by the 
impinging diffuse ionizing field, formed in the surrounding nebula. As we have
 done for the heads of the cometary globules, we compare the emission of the
 cometary tails with the predictions from our model in order to quantitatively
 determine the value of the diffuse ionizing photon flux at the positions of 
the globules.

Knowing the direct stellar radiation and the diffuse flux at different 
distances from the central star (i.e., at the positions of the cometary 
globules), we can then calculate the diffuse-to-direct ionizing flux ratio 
($F_d/F_*$) as a function of distance from the central star. We compare 
the data with simple models for the diffuse ionizing field 
(Henney 2000) and conclude that the diffuse field is intermediate in 
magnitude between that expected from a filled and shell-like geometry, 
which may be due to the disk-like nature of the Helix nebula.

\acknowledgments
L. L\'opez-Mart\'\i n is in grateful receipt of a graduate scholarship from 
DGEP-UNAM (M\'exico). L. L\'opez-Mart\'\i n,  A. C. Raga and J. Cant\'o 
acknowledge support from the CONACyT grants 27546-E and 32753-E.
 W. J. Henney acknowledges
 support from CONACyT grant 27570E and 27546-E, and DGAPA grant IN128698.
We thank R. C. O'Dell for useful comments and suggestions. We also thank
an anonymous referee for helpful comments.


\begin{thebibliography}{}
\bibitem[Bertoldi & Mckee]{Ber90} Bertoldi, F. \& McKee, C. F. 1990, ApJ, 354, 529
\bibitem[Burkert \& O'Dell 1998]{bur98} Burkert, A. \& O'Dell, C. R. 1998, ApJ, 503, 792
\bibitem[Cahn \& Kaler]{CK71}Cahn, J. H.\& Kaler, J. B. 1971, ApJS, 22, 319
\bibitem[Cahn \& Kaler \& Stanghellini]{CKS92} Cahn, J. H., Kaler, J. B. \& Stanghellini, L. 1992, A\&AS, 94, 399
\bibitem[Cant\'o \& Raga \& Steffen \& Shapiro]{Can98} Cant\'o, J., Raga, A. C., Steffen, W. \& Shapiro, P. 1998, ApJ, 502, 695
\bibitem[Capriotti]{Cap73} Capriotti, E. R. 1973, ApJ, 179, 495
\bibitem[Daub]{D82} Daub, C. T. 1982, ApJ, 253, 679
\bibitem[Dyson]{Dys93} Dayson, J. E., Harquist, T. W. \& Biro, S. 1993, MNRAS, 261, 430
\bibitem [Gorny \& Stasinska \& Tylenda]{Gorny97} G\'orny, S.K., Stasi\'nska, G. \& Tylenda, R. 1997, A\&A, 318, 256
\bibitem[Harris  \& Dahn  \& Monet  \& Pier] {H97} Harris, H.C., Dahn, C.C., Monet, D.G.  \& Pier, J.R. 1997, in IAU Symp. 180, Planetary Nebulae, ed. H. Habing (Dordretch : Reidel), 40
\bibitem[Henney]{Hen00} Henney, W. J. 2000, in preparation 
\bibitem[Henney \& Arthur]{Hen98} Henney, W. J. \& Arthur, S.J. 1998, AJ, 116, 322
\bibitem[Henry \& Kwitter \& Dufour]{Henry99} Henry, R.B.C., Kwitter, K.B. \& Dufour, R.J. 1999, ApJ, 517, 782
\bibitem[Huggins  \& Bachiller  \& Cox  \& Forveille] {H92} Huggins, P.J., Bachiller, R., Cox, P.  \& Forveille, T. 1992, ApJ, 401, 43
\bibitem[Johnstone98]{John98} Johnstone, D., Hollenbach, D. \& Bally, J. 1998, ApJ, 499, 758
\bibitem[Meaburn \& CLayton \& Bryce \& Walsh \& Holloway \& Steffen]{Meaburn92} Meaburn, J., Clayton, C. A., Bryce, M., Walsh, J. R., Holloway, A. J. \& Steffen, W. 1998, MNRAS, 294, 201
\bibitem[Meaburn \& CLayton \& Bryce \& Walsh]{Mea96} Meaburn, J., Clayton, C. A., Bryce, M., Walsh, J. R. 1996, MNRAS, 281, L57
\bibitem[Meaburn \& Walsh \& Clegg \& Walton \& Taylor \& Berry]{Mea92} Meaburn, J., Walsh, J. R., Clegg, R. E. S., Walton, N. A., Taylor, D. \& Berry, D. S. 1992, MNRAS, 255, 177
\bibitem[Meaburn \& White] {M82} Meaburn, J. \& White, N.J. 1982, Ap\&SS, 82, 423
\bibitem[Mellema \& Raga \& Cant\'o \& Lundqvist \& Balick \& Steffen \& Noriega-Crespo]{Mell98} Mellema, G., Raga, A. C., Cant\'o, J., Lundqvist, P., Balick, B., Steffen, W. \& Noriega-Crespo, A. 1998, A\&A, 331, 335
\bibitem[O'Dell2]{O982} O'Dell, C. R. 1998, AJ, 115, 263
\bibitem[O'Dell]{O98} O'Dell, C.R. 1998, AJ, 116, 1346
\bibitem[O'Dell99]{Odell99} O'Dell, C. R. \& Doi, T. 1999, PASP, 111, 1316
\bibitem[O'Dell \& Burkert]{OB97} O'Dell, C. R. \& Burkert, A. 1997, in IAU Sym
p. 180, Planetary Nebulae, ed. H. Habing (Dordretch : Reidel), 332
\bibitem[O'Dell \& Handron]{OH96} O'Dell, C. R. \& Handron, D. 1996, AJ, 96, 23
\bibitem[osterbrock]{Oster89} Osterbrock, D. E. 1989, Astrophysics of Gaseous Nebulae and Active Galactic Nuclei (Mill Valley : University Science Books)
\bibitem [Rauch]{Rauch97} Rauch, T. 1997, A\&A, 320, 237
\bibitem[Vorontsov-Velyaminov]{Vor68} Vorontsov-Velyaminov, B. A. 1968, in Planetary Nebulae: IAU Symposium 34, eds. D. E. Osterbrock \& C. R. O'Dell (Dordrecht:Reidel) 256
\bibitem[Walder]{Wal98} Walder, R., Folini, D. 1998, A\&A, 330, L21
\bibitem[Young \& Cox \& Huggins \&  Forveille  \& Bachiller]{Young98} Young, K., Cox, P., Huggins, P. J., Forveille, T. \& Bachiller, R. 1997, ApJ, 482, L101
\bibitem[Young \& Cox \& Huggins \&  Forveille  \& Bachiller]{Young99} Young, K., Cox, P., Huggins, P. J., Forveille, T. \& Bachiller, R. 1999, ApJ, 522, 387


\end{thebibliography}
\end{document}